\documentclass[sigconf]{acmart}

\setcopyright{none}
\renewcommand\footnotetextcopyrightpermission[1]{}
\acmConference{}{}{}
\acmBooktitle{}
\acmYear{}
\copyrightyear{}
\acmDOI{}
\acmISBN{}

\AtBeginDocument{%
  }

\usepackage{xurl}
\usepackage{booktabs}
\usepackage{amsmath}
\usepackage{xcolor}
\usepackage{dsfont}
\usepackage{tabularx}
\usepackage{graphicx}  
\usepackage{hyperref} 

\usepackage{tcolorbox}
\tcbuselibrary{breakable}
\usepackage{enumitem}
\usepackage{seqsplit}
\newtcolorbox{findingbox}{%
  breakable,
  colback=gray!5, colframe=gray!55,
  fontupper=\small,
  boxrule=0.5pt, arc=2pt,
  left=5pt, right=5pt, top=4pt, bottom=4pt,
}
\newtcolorbox{casebox}[1]{%
  breakable,
  colback=cyan!4, colframe=cyan!45!black,
  coltitle=black, colbacktitle=cyan!15,
  fonttitle=\bfseries\small, fontupper=\small,
  boxrule=0.5pt, arc=2pt,
  left=5pt, right=5pt, top=4pt, bottom=4pt,
  before skip=6pt, after skip=6pt,
  title={#1}%
}
\newtcolorbox{promptbox}[1]{%
  colback=black!2, colframe=black!45,
  coltitle=white, colbacktitle=blue!35!black!65,
  fonttitle=\bfseries\small, fontupper=\small,
  before upper=\raggedright,
  boxrule=0.5pt, arc=2pt,
  left=5pt, right=5pt, top=4pt, bottom=4pt,
  title={#1}%
}
\newcommand{\wbadd}[1][\textcolor{black}]{#1}

\usepackage{orcidlink}
\usepackage[misc]{ifsym}

\begin{document}
\pagestyle{plain}

\title{Skills That Don't Exist: A Large-Scale Study of Hallucinated Skill Recommendation in LLM Agents}
\thanks{\textsuperscript{\Letter} Wenbo Guo and Feng Dong are corresponding authors.}

\author{Weifeng Yuan}
\orcid{0009-0000-4910-3041}
\affiliation{%
  \institution{Huazhong University of Science and Technology}
  \city{Wuhan}
  \country{China}
}
\email{cougarsecu@gmail.com}

\author{Wenbo Guo\textsuperscript{\Letter}}
\orcid{0000-0001-6655-8179}
\affiliation{%
  \institution{Nanyang Technological University}
  \country{Singapore}
}
\email{honywenair@gmail.com}

\author{Feng Dong\textsuperscript{\Letter}}
\orcid{0000-0001-7091-2169}
\affiliation{%
  \institution{Huazhong University of Science and Technology}
  \city{Wuhan}
  \country{China}
}
\email{dongfeng@hust.edu.cn}

\author{Haoyu Wang}
\orcid{0000-0003-1100-8633}
\affiliation{%
  \institution{Huazhong University of Science and Technology}
  \city{Wuhan}
  \country{China}
}
\email{haoyuwang@hust.edu.cn}

\author{Yang Liu}
\orcid{0000-0001-7300-9215}
\affiliation{%
  \institution{Nanyang Technological University}
  \country{Singapore}
}
\email{yangliu@ntu.edu.sg}

\begin{abstract}


LLM agents dynamically acquire new capabilities by downloading \emph{skills} from open registries. Instead of browsing these catalogs manually, developers typically ask the agent to recommend and install a skill for them. This convenience hides a risk: agents frequently invent names for skills that do not exist, confidently returning names that appear in no registry. We term this flaw \emph{skill name hallucination}. A fake name may seem harmless, but it acts as an open door for supply-chain attacks. Because registries rarely verify publishers, an adversary can prompt the agent, collect the fake names it returns, and pre-register malicious skills under them, then wait for a victim to install the payload.


To uncover the severity of this threat, we conducted the first large-scale measurement of skill name hallucination. We evaluated 15,000 prompts across 12 configurations (4 standalone LLMs and 8 agents). To ensure absolute accuracy, we conservatively counted a name as hallucinated only if it was missing from all live registries and GitHub. The results reveal a systemic vulnerability, as every single configuration hallucinates. Hallucination rates average 36.0\% for standalone LLMs and 36.9\% for agents, rising to 43.1\% when answering real-world developer questions. In total, the systems generated 5,669 distinct hallucinated names. Crucially, these names are not random noise. Agents predictably repeat the same fake names across different prompts and models, giving attackers highly reliable targets to hijack.

Finally, we tested four model-level defenses and found a severe conflict between security and usability. The strongest defense, retrieval grounding, successfully slashed the hallucination rate from 40.8\% down to 3.2\%. However, it crippled the agent's usefulness. Even the best-defended system could only recommend the correct skill about one out of every six times. Ultimately, skill name hallucination is a highly exploitable vulnerability requiring minimal attacker effort. Fixing it cannot rely on prompt engineering or model tuning alone. It demands ecosystem-wide structural changes, specifically registry-level name reservations and verified recommendation pipelines.

  
\end{abstract}

\ccsdesc[500]{Computing methodologies~Artificial intelligence}
\ccsdesc[500]{Software and its engineering~Software libraries and repositories}
\ccsdesc[300]{Security and privacy~Malware and its mitigation}


\maketitle

\section{Introduction}
\label{sec:introduction}

To add a new capability to an AI coding agent, developers used to browse a
registry. Most of the time they just ask the agent which skill to use, and install whatever it names. A \emph{skill} is a small bundle of natural-language instructions, kept in a \texttt{SKILL.md} file, that teaches an agent to do one task. The Agent Skills
ecosystem is new, but it has grown quickly. Anthropic published the Agent Skills standard in December 2025, and OpenAI adopted it within
weeks~\cite{anthropic_skills_2025,anthropic_release_notes,anthropic_agent_skills,openai_skills_repo_2026,openai_codex_skills,anthropic_skills_overview,claude_agent_sdk_skills,openai_api_skills,github_copilot_skills,vscode_agent_skills,antigravity_skills,cursor_skills,opencode_skills}. Analysts expect
agent-enabled enterprise applications to rise from under 5\% in 2025 to 40\% the
following year~\cite{gartner_agents_2025}. The registries kept pace, and SkillsMP
alone lists more than 1.7 million skills~\cite{skillsmp}. Nobody can read
through a catalog that large, so the agent decides and the user installs
what it suggests. This habit is visible online. On Stack Overflow, Reddit,
and Hacker News, developers routinely ask for skill and tool recommendations~\cite{latendresse2024chatgpt,sandoval2023lost,klemmer2024aiassistant}.

This delegation rests on a quiet assumption. The skill the agent names should
exist. Often it does not. Language models \emph{hallucinate}, and they state
invented facts with full confidence~\cite{ji2023_hallucination_survey}. Inside an
agent, this shows up as a recommended skill whose name matches no entry in any
registry. We call this a \emph{skill name hallucination}. This is not a rare error.
On real user questions it is the common case, and it persists even when the agent
has a web-search tool and is told to verify the name first. These hallucinated
skill names are not random. The model falls back on capabilities and tools it
already knows. Some are generic capability phrases, such as
\texttt{file-system-operations} or \texttt{text-editor-tool}. Others borrow names
of familiar tools and products, such as \texttt{visual-studio-code} or
\texttt{ublock-origin}. A web search finds results for all of these, because they
name real capabilities or tools. The search confirms that the name means
something, but not that any published skill has that name. Web search runs the
wrong check.

A hallucinated name is harmless only while nothing exists behind it. An attacker
removes that condition in three steps. The attacker first queries public agents
and records the names they invent again and again. The attacker then registers one
of those names on an open registry, with malicious instructions hidden in the
\texttt{SKILL.md}. This step is cheap, because these registries barely check who
publishes. Later, a victim who asks a similar question receives the same name and
installs it. Installation is not a real barrier. Many agents auto-approve it, and
even when a prompt appears, users tend to approve a recommendation from their own
agent~\cite{perry2023users}. The same agent that recommends the skill is the one
that later runs it (Section~\ref{sec:threat}). No existing supply-chain or
injection defense stops the attack, because a hallucinated name gives a defender
nothing to work with. There is no real name to match, and the payload can hide in
plain language rather than code~\cite{guo2026malskillbench}. This is what separates the threat from
package squatting~\cite{spracklen2025we}, and Section~\ref{sec:rq3} measures
the resulting detection failure.

To measure how real this threat is, we run the first systematic study of skill
name hallucination. We send 15{,}000 prompts to 12 evaluated configurations and check every name
they recommend against the live registries. The prompts are of two kinds, community prompts
drawn from developer forums and skill description prompts built from real published skills. The 12
evaluated configurations comprise 4 standalone LLM configurations and 8 agent configurations, most of them
given a web-search tool. We organize the study around four questions that
follow the attacker's path:

\begin{itemize}[leftmargin=*]
  \item \textbf{RQ1 (Prevalence).} How often do agents hallucinate skill names, and how robust is that rate to the model, its temperature, its size, and the way users ask?
  \item \textbf{RQ2 (Targetability).} How much attack coverage is possible by registering recurring hallucinated skill names?
  \item \textbf{RQ3 (Detectability).} Why are hallucinated skill names hard to detect by the model itself and by edit-distance filters?
  \item \textbf{RQ4 (Mitigation).} How well do retrieval-augmented generation (RAG), self-refinement, and cross-model voting reduce hallucination, and at what cost to the agent's usefulness?
\end{itemize}

Every evaluated configuration hallucinates, at rates from 6.5\% to 62.0\%. The surface is
widest where it matters most. On synthetic prompts, where a correct skill is known
to exist, the mean rate is 29.4\%, but on real developer questions it rises to
43.1\%. No model, temperature, size, or phrasing removes the problem, and a
web-search tool does not either. The names are also easy to target. In the current
live-validated persistence run, Claude~Sonnet~4.6 repeats its dominant hallucinated skill
name 7.8 times in 10. Most names are specific to one configuration, but 410 are shared by
two or more configurations, and these shared names account for 15.3\% of all hallucinated
recommendation events. The same naming prior also reaches neighboring software
ecosystems, where 851 hallucinated skill names are real PyPI or npm packages.

The names are also hard to detect. A model asked to verify a name accepts real and hallucinated skill names at nearly the same rate, yielding 51\% balanced accuracy, no better than a coin flip. They sit a median of six edits from any real skill, far beyond the reach of typosquatting filters. Defenses narrow the surface but do not close it. RAG cuts confirmed hallucination by 37.6 percentage points on average, from 40.8\% to 3.2\%, while self-refinement proves far less effective from 40.8\% to 33.2\%. Even the best-defended agent names the right skill only about one time in six, so it is safe mostly because it returns no output. The ecosystem stays hard to use safely.

We make the following contributions:
\begin{itemize}[leftmargin=*]
    \item \textbf{A new attack on the agent skill supply chain.} We identify skill
    name hallucination as the entry point of a skill name squatting attack, in which an
    agent's own persistent hallucinations let an attacker pre-register the names
    that the agent will later recommend, and we show why this falls outside the
    reach of existing supply-chain and prompt-injection defenses
    (Section~\ref{sec:threat}).

    \item \textbf{A measurement of the attack surface and its targetability.}
    Across 15{,}000 prompts and 12 evaluated configurations, we measure how prevalent, how
    reusable, and how stealthy hallucinated skill names are, including a new
    skill name squatting coverage analysis and a cross-ecosystem overlap between skills and
    Python and npm packages (Section~\ref{sec:evaluation}).

    \item \textbf{An evaluation of defenses and their cost.} We test four
    defenses and report both their effect on the attack surface and
    their cost to the agent's usefulness (Section~\ref{sec:mitigation}).
\end{itemize}

\section{Background}

\subsection{Skill Name Hallucination}

\wbadd{Agents load skills progressively. The agent first sees only a skill's \texttt{name} and \texttt{description}, and reads the full \texttt{SKILL.md} instructions only after it selects the skill. The name is therefore both a
discovery signal and a security-sensitive input.}
\wbadd{Skill selection fails when the agent names a skill that does not exist. We call such a recommendation a \emph{skill name hallucination}. The failure is similar to a known one in code generation, where models recommend packages that do not exist on PyPI or npm~\cite{pearce2025asleep, spracklen2025we}. }

\subsection{Threat Model}
\label{sec:threat}

\begin{figure}[htbp]
    \centering
\includegraphics[width=1\linewidth]{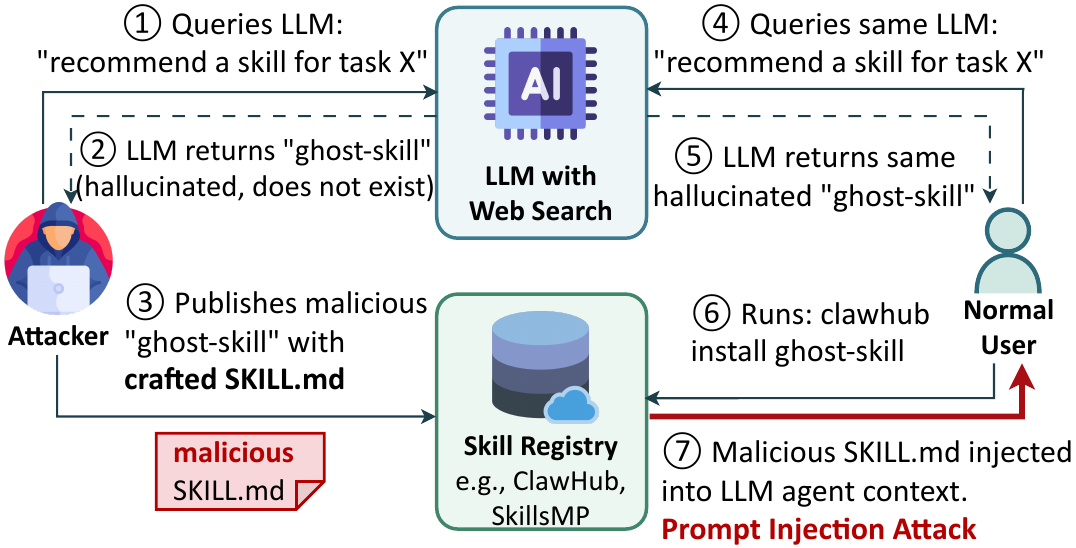}
    \caption{Overview of the skill hallucination and prompt injection attack.}
    \label{fig:threat_model}
\end{figure}

\textbf{Participants.}
We model four parties: the \emph{agent}, which recommends and later loads a
skill; the \emph{registry}, such as ClawHub or SkillsMP, which accepts
community skills; the \emph{victim}, who installs the returned name; and the
\emph{attacker}, an ordinary user of the same public agents and registries.
The attack relies on misplaced trust: the victim trusts the agent, the registry
trusts publishers, and the agent cannot distinguish a real skill from one it
invented (Section~\ref{sec:rq3}). Unlike package hallucination, one agent can
invent, recommend, and later execute the installed skill.

\textbf{Attack flow.}
The attack has three steps (Figure~\ref{fig:threat_model}). First, the attacker
queries public agents and records skill names that recur. Second, the attacker
publishes a skill under one such name, with malicious instructions or scripts in
its \texttt{SKILL.md} package. This requires little privilege: ClawHub accepts
week-old accounts, and SkillsMP indexes public repositories. Third, a later
victim receives the same hallucinated name and installs it. Even when
installation is not automatic, prompts often show only the name, not the hidden
instructions.

\section{Experiment Design}
\label{sec:experiment}

To measure skill name hallucination across models and agents, we use a three-stage pipeline. (1)~\emph{Prompt Construction} builds a dataset of skill-seeking prompts. (2)~\emph{Recommendation Generation} sends each prompt to every model and agent and records the skill names they return. (3)~\emph{Skill Name Validation} checks each returned name against the live registries and counts any name no registry contains as a hallucination.

\subsection{Prompt Construction}
\label{sec:prompt_construction}

Our dataset has 15{,}000 prompts of two types, each measuring a different aspect of hallucination. \emph{Community Prompts} are real questions drawn from developer forums. They measure how often an agent hallucinates on the questions users actually
ask. Many of these questions have no matching skill, so the model must either state that none exists or invent a name. \emph{Skill Description Prompts} are built from the \texttt{description} field of a real published skill's \texttt{SKILL.md}, the short text that states what the skill does. Because each prompt comes from a real skill, a correct skill is known to exist for every one. These prompts measure whether a model can output a correct skill name it should be able to find. Together, the two types separate hallucination caused by a missing skill from hallucination that remains even when the right skill exists.

\subsubsection{Community Prompts}
\label{sec:community_dataset}

To make the hallucination rate reflect a real attack surface rather than an artificial one, we collect prompts that match the questions real users ask. We mine real questions from three developer communities: Stack Exchange technical sites, AI and developer subreddits, and Hacker News (retrieved through the Algolia API). In each source, we search
for skill-seeking questions with queries such as ``what tool can I use,'' ``best
plugin,'' and similar skill recommendation phrases.

\wbadd{Most forum posts are irrelevant to our task, so before building prompts we score each post with a set of relevance rules. The rules use the post title, body, and source community. They add evidence for a software skill question when a post contains a target term (\emph{skill}, \emph{tool}, \emph{plugin}, \emph{extension}, \emph{mcp}, \emph{agent}), software or AI context (e.g., \emph{github}, \emph{codex}, \emph{claude}, \emph{llm}), a technical community (e.g., Stack Overflow, \emph{r/vscode}), or a question form. They subtract evidence when a post uses \emph{skill} in the human sense (e.g., \emph{soft skills}, \emph{cooking skills}), comes from a nontechnical community (e.g., \emph{r/AskReddit}), or mentions a target term without technical context, which usually refers to a physical tool or human ability. We keep only posts whose score exceeds a threshold, so genuine software skill questions pass while irrelevant posts fall below the cutoff. The result is 5{,}000 community prompts covering skill name inquiries (2{,}395), skill recommendations (1{,}842), and usage descriptions (763), drawn from Stack Exchange (2{,}484), Reddit (1{,}878), and Hacker News (638).}

\subsubsection{Skill Description Prompts}
\label{sec:registry_dataset}

\wbadd{Community prompts show how often agents hallucinate in the wild, but for many of them no real skill fits. We therefore cannot tell whether a hallucination comes from a missing skill or happens even when a correct skill exists. Skill description prompts resolve this ambiguity, because each one is built from a real published skill, so a correct answer is known to exist. We take the 10{,}000 most popular skills on \textit{SkillsMP} and turn each skill's description field into one prompt that asks the model to return the matching skill's name. Because each prompt comes from a real skill, a correct answer always exists, so any hallucination here cannot be attributed to a missing skill.}


\subsection{Recommendation Generation}
\label{sec:model_selection}

\wbadd{We evaluate skill name hallucination on two kinds of system, standalone LLMs and LLM-powered agents. The main design challenge is the knowledge cutoff. Anthropic released the Agent Skills standard on December 18, 2025~\cite{anthropic_skills_2025}, after the training cutoff of most models we test. These models could then hallucinate because they never saw a real skill during training, not because of any deeper tendency. To rule out this explanation, we give every pre-cutoff model and agent a web-search tool~\cite{openai_function_calling_2023}, require it to search before it recommends a skill, and confirm that search works for each one. The one exception is Claude Opus~4.7 (cutoff January 2026), which already knows the standard. We run it without search to approximate the best case for training-time knowledge.}


\wbadd{Table~\ref{tab:models} lists the models and agents we evaluate. We use four standalone LLM configurations and eight agent configurations spanning four agent frameworks. Claude Code and Codex search the web natively, while OpenClaw and OpenCode use a self-hosted SearXNG instance through MCP. We run OpenCode with five backends (GPT-5.4-mini and four open-source models). Together, the 4 standalone LLM configurations and 8 agent configurations give the 12 configurations we report throughout the paper. Before running the skill prompts, we checked that web search worked for every search-enabled system. We asked each one to fetch a current-news item on the test day. This check only confirmed that search was reachable, and we excluded it from the hallucination measurements.}

\begin{table}[htbp]
\centering
\caption{Evaluated configurations}
\label{tab:models}
\scriptsize
\setlength{\tabcolsep}{4pt}
\renewcommand{\arraystretch}{1.15}
\begin{tabularx}{\columnwidth}{@{}llXll@{}}
\toprule
\textbf{System} & \textbf{Type} & \textbf{Model} & \textbf{Knowledge} & \textbf{Web Search} \\
\midrule
GLM-4.5-Air       & LLM & glm-4.5-air        & Pre-release & Enabled \\
GPT-5.4-mini      & LLM & gpt-5.4-mini       & Pre-release & Enabled \\
Claude Sonnet 4.6 & LLM & sonnet-4-6  & Pre-release & Enabled \\
Claude Opus 4.7   & LLM & opus-4-7    & Jan 2026    & Disabled \\
\midrule
Claude Code & Agent & sonnet-4-6 & Pre-release & Enabled \\
Codex       & Agent & gpt-5.3-codex     & Pre-release & Enabled \\
OpenClaw    & Agent & gpt-5.4-mini      & Pre-release & SearXNG (MCP) \\
OpenCode    & Agent & gpt-5.4-mini & Pre-release & SearXNG (MCP) \\
OpenCode    & Agent & 4 Ollama Models & Pre-release & SearXNG (MCP) \\
\bottomrule
\end{tabularx}
\parbox{\linewidth}{\scriptsize The OpenCode row with four Ollama models expands to four agent configurations: Ministral-3~8B, Granite~3.3~8B, Qwen~3.5~9B, and Nemotron-Cascade-2~30B. Later figures shorten these names to Min3-8B, Gra3.3-8B, Qwen3.5-9B, and Nem2-30B.}
\end{table}

\begin{figure}[htbp]
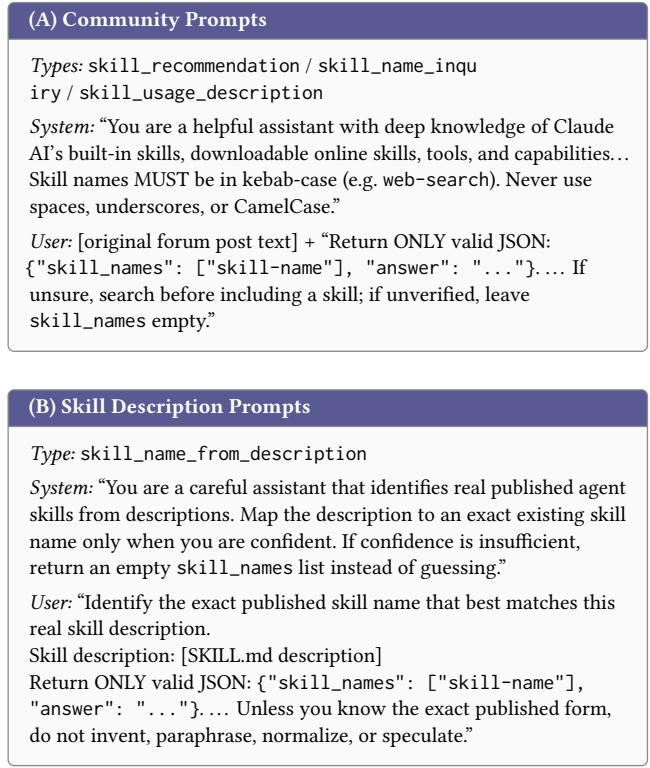

\centering
\small
\begin{promptbox}{(A) Community Prompts}
\textit{Types:} \texttt{skill\_recommendation} / \texttt{skill\_name\_inqu\\iry} / \texttt{skill\_usage\_description}\\[3pt]
\textit{System:} ``You are a helpful assistant with deep knowledge of Claude AI's built-in skills, downloadable online skills, tools, and capabilities\ldots{} Skill names MUST be in kebab-case (e.g.\ \texttt{web-search}). Never use spaces, underscores, or CamelCase.''\\[3pt]
\textit{User:} \textnormal{[original forum post text]} + ``Return ONLY valid JSON: \texttt{\{"skill\_names": ["skill-name"], "answer": "..."\}}. \ldots{} If unsure, search before including a skill; if unverified, leave \texttt{skill\_names} empty.''
\end{promptbox}

\vspace{4pt}
\begin{promptbox}{(B) Skill Description Prompts}
\textit{Type:} \texttt{skill\_name\_from\_description}\\[3pt]
\textit{System:} ``You are a careful assistant that identifies real published agent skills from descriptions. Map the description to an exact existing skill name only when you are confident. If confidence is insufficient, return an empty \texttt{skill\_names} list instead of guessing.''\\[3pt]
\textit{User:} ``Identify the exact published skill name that best matches this real skill description.

Skill description: \textnormal{[SKILL.md description]}

Return ONLY valid JSON: \texttt{\{"skill\_names": ["skill-name"], "answer": "..."\}}. \ldots{} Unless you know the exact published form, do not invent, paraphrase, normalize, or speculate.''
\end{promptbox}
\caption{Prompt templates.}
\label{fig:prompt-templates}
\end{figure}

Figure~\ref{fig:prompt-templates} shows the two prompt templates.
For \emph{community prompts}, \wbadd{the system message instructs the model to act as a skill recommender} and requires kebab-case names. The user message adds the forum post, the required output format, and an instruction to search the web first.
For \emph{skill description prompts}, the system message asks the model to identify the one real skill that matches, and to return an empty \texttt{skill\_names} list rather than guess when it is unsure.
Both templates require a JSON reply of the form \texttt{\{"skill\_names": [...], "answer": "..."\}}.
We cap each reply at 1{,}200~tokens and leave all sampling parameters at provider defaults, except in RQ2, where we vary them on purpose.

\subsection{Skill Name Validation}
\label{sec:validation}

\wbadd{We treat a recommended skill as a hallucination when it exists on no skill-hosting registry and in no open-source repository. To verify whether a skill is real, we use a two-stage validation. \textbf{\textit{Registry verification.}} We match the name against a broad reference index built from multiple public sources of real skill names, including the SkillsMP and ClawHub registries, the curated awesome-openclaw list, Claude and Codex skill repositories, historical \texttt{SKILL.md} records, and our prior registry lookups. SkillsMP in particular is built by periodically harvesting skill from public GitHub repositories and now indexes over 1.7~million skills, so a name missing even from this large catalog is already very likely fake. \textbf{\textit{Expanded verification.}} For every name the registries do not resolve, we expand the search to GitHub. We query GitHub Code Search with \texttt{"name: <canonical-name>" filename:SKILL.md}, fetch the candidate
\texttt{SKILL.md} files, and accept a match only when a candidate's front-matter \texttt{name:} field canonicalizes to the generated name. This stage is slow and costly, but it ensures that a real skill too new to appear in the registries is never miscounted. We label a name a hallucination only when both stages fail.}



We take the names from the \texttt{skill\_names} field of each response and match them against this validation pipeline. We first normalize both generated and
reference names by trimming wrappers and markup, splitting camel case, converting \texttt{\&} to \texttt{and}, replacing non-alphanumeric runs with hyphens, and
lowercasing. \wbadd{A name counts as real on an exact match, a unique canonical-alias match, or a matching \texttt{name:} field in a \texttt{SKILL.md} found by either stage. We discard clear extraction artifacts, such as URLs, paths, schema tokens, trace IDs, and sentence-like strings. None of these can be a skill name, which is a short kebab-case identifier, so they are noise from parsing the response rather than genuine recommendations. We exclude them instead of counting them as hallucinations. We cache all lookups to avoid repeated searches.}

\wbadd{Our primary metric is the \emph{skill name hallucination rate}. We compute this metric at the unique-name level, not over repeated recommendation events. Let $R$ be the set of distinct canonical names a system recommends, after we discard the extraction artifacts above. For each name $r \in R$, the indicator $\mathds{1}[,r \text{ is hallucinated},]$ equals $1$ when $r$ fails both verification stages and $0$ otherwise. The hallucination rate is the share of these unique recommended names that are hallucinated:}

\begin{equation}
\label{eq:hallucination_rate}
\text{Hallucination Rate} =
\frac{\sum_{r \in R}\mathds{1}[,r \text{ is hallucinated},]}{|R|}
\end{equation}

\wbadd{A higher rate means a wider attack surface, because more of an agent's
recommendations point to skills that do not exist. On the skill description prompts, where a correct skill is known to exist, we additionally report \emph{recall}, the fraction of prompts for which the model returns the correct name.}


\section{Empirical Results}
\label{sec:evaluation}

\wbadd{We investigate four research questions, each on a distinct dimension of skill name hallucination. RQ1 measures prevalence, how often agents hallucinate and whether the rate holds across model family, temperature, size, and prompt domain. RQ2 measures targetability, how often hallucinated names repeat, how widely they are shared across configurations, how often they overlap with real software packages, and how few names cover most cases. RQ3 measures detectability, whether self-auditing or lexical distance to real skill names can identify hallucinated names. RQ4 measures mitigation, how well RAG, self-refinement, and cross-model voting reduce hallucination.}

\subsection{RQ1: Prevalence}
\label{sec:rq1}

\wbadd{To systematically measure the prevalence of this attack surface, we evaluated all 12 configurations on the full 15{,}000-prompt corpus. Every recommended name was validated using the two-stage pipeline from Section~\ref{sec:validation}. To avoid duplicate counting, hallucination rates are computed over unique skill names rather than repeated recommendation events. \autoref{tab:rq1_summary} reports these statistics, separating name-level metrics and prompt-level outcomes to illustrate the varying attack surfaces under different external grounding methods. The results demonstrate that hallucination is ubiquitous, with rates ranging from 6.5\% for GPT-5.4-mini to 62.0\% for OpenCode with Ministral-3~8B, averaging 36.0\% across standalone LLMs and 36.9\% across agents.}

\begin{table}[htbp]
  \centering
  \caption{Per-configuration hallucination statistics}
  \label{tab:rq1_summary}
  \setlength{\tabcolsep}{3.5pt}
  \resizebox{\linewidth}{!}{
  \begin{tabular}{llcccccc}
    \toprule
    \textbf{Model} & \textbf{Search} & \textbf{Unique} & \textbf{Matched} & \textbf{Halluc.} & \textbf{Other} & \textbf{H. Rate} & \textbf{No output} \\
    \midrule
    \multicolumn{8}{l}{\textit{LLM direct API}} \\[1pt]
    Claude~Sonnet~4.6              & Enabled  & 6{,}039 & 2{,}970 & 2{,}982 & 87 & 49.4\% & 26.6\% \\
    GPT-5.4-mini                   & Enabled  & 5{,}602 & 5{,}207 & 363 & 32 & 6.5\% & 52.5\% \\
    GLM-4.5-Air                    & Enabled  & 5{,}771 & 2{,}695 & 2{,}965 & 111 & 51.4\% & 46.0\% \\
    Claude~Opus~4.7                & Disabled & 2{,}051 & 1{,}292 & 753 & 6 & 36.7\% & 56.8\% \\
    \midrule
    \multicolumn{8}{l}{\textit{Coder Agent}} \\[1pt]
    Claude~Code~Sonnet~4.6         & Enabled  & 234 & 153 & 81 & 0 & 34.6\% & 90.9\% \\
    OpenClaw~GPT-5.4-mini          & SearXNG  & 4{,}677 & 3{,}355 & 1{,}193 & 129 & 25.5\% & 56.1\% \\
    OpenCode~GPT-5.4-mini          & SearXNG  & 1{,}231 & 898 & 287 & 46 & 23.3\% & 92.2\% \\
    Codex~GPT-5.3-Codex            & Enabled  & 3{,}770 & 2{,}601 & 1{,}100 & 69 & 29.2\% & 76.2\% \\
    OpenCode~Ministral-3~8B        & SearXNG  & 4{,}830 & 1{,}276 & 2{,}994 & 560 & 62.0\% & 81.8\% \\
    OpenCode~Granite~3.3~8B        & SearXNG  & 4{,}874 & 2{,}027 & 2{,}693 & 154 & 55.3\% & 69.0\% \\
    OpenCode~Qwen~3.5~9B           & SearXNG  & 4{,}551 & 2{,}716 & 1{,}719 & 116 & 37.8\% & 80.0\% \\
    OpenCode~Nemotron-Cascade-2~30B & SearXNG & 4{,}389 & 3{,}094 & 1{,}208 & 87 & 27.5\% & 69.3\% \\
    \bottomrule
  \end{tabular}
  }
  \vspace{0.15cm}
  \parbox{\linewidth}{\scriptsize
\emph{Unique} equals \emph{Matched}, \emph{Halluc.}, and \emph{Other}, which represent verified real skills, unverified hallucinations, and extraction artifacts, respectively. \emph{H. Rate} is the ratio of \emph{Halluc.} to \emph{Unique}. \emph{No output} indicates the fraction of prompts yielding no skill recommendation.}
\end{table}

\wbadd{Rather than generating arbitrary noise, the models predominantly output semantically plausible but non-existent skill identifiers. Qualitative analysis reveals two dominant categories. The first comprises abstract capability descriptions (e.g., \texttt{file-system-operations} or \texttt{text-editor-tool}), where the model attempts to describe the required action instead of recalling a verified skill. The second involves adopting the names of real-world software or ecosystem objects that have not been officially registered as skills (e.g., \texttt{visual-studio-code} or \texttt{ublock-origin}).}

External grounding via web search fails to mitigate this vulnerability. Despite enabling three of the four standalone LLMs with built-in search tools and explicitly prompting them to verify names prior to output, hallucination rates remain critically high (e.g., 49.4\% for Claude~Sonnet~4.6 and 51.4\% for GLM-4.5-Air). This failure stems from the semantic plausibility of the hallucinated names, because searching for \texttt{file-system-operations} returns abundant relevant results, thereby tricking the agent into confirming the concept's existence rather than checking registry status.

Furthermore, conservative output generation artificially deflates hallucination rates without addressing the underlying threat. For instance, GPT-5.4-mini achieves a deceptively low rate of 6.5\% primarily because it returns no skill on 52.5\% of the prompts. This behavior is more extreme in agent configurations, with Claude~Code returning no output for 90.9\% of prompts and yielding the fewest names overall, yet 34.6\% of its outputs are still hallucinated. Consequently, higher no-output rates simply limit output volume, leaving the output skill names highly vulnerable to hallucination.


\begin{findingbox}{Finding 1. Skill name hallucination is ubiquitous and persists despite basic safeguards.}
All 12 configurations hallucinate skill names, with rates from 6.5\% to 62.0\%.
The root cause is that models often turn a capability into a name, such as
\texttt{file-system-operations} or \texttt{text-editor-tool}, instead of recalling
a published skill. Furthermore, two system behaviors fail to fix this. Web search can confirm that such phrases
exist, but not that they are valid skill names. Finally, more conservative output reduces the volume of recommended names, but the names that remain are often hallucinated.
\end{findingbox}

The evaluation of Claude~Opus~4.7 provides insights into whether recent training data inherently neutralizes the attack surface. As its January 2026 knowledge cutoff postdates the December 2025 Agent Skills release~\cite{opus47}, this configuration isolates the impact of training-time knowledge updates. Operating without web search, the model adopts a highly restrictive generation strategy, returning no skills for 56.8\% of prompts. Although its hallucination rate drops to 36.7\%, falling below that of Claude~Sonnet~4.6, it remains higher than several search-enabled agents. For example, OpenCode with GPT-5.4-mini and OpenClaw with GPT-5.4-mini achieve lower rates of 23.3\%, 25.5\%, respectively. Thus, updated training corpora reduce output volume but do not remove the attack surface.

\begin{figure}[htbp]
  \centering
  \includegraphics[width=\linewidth]{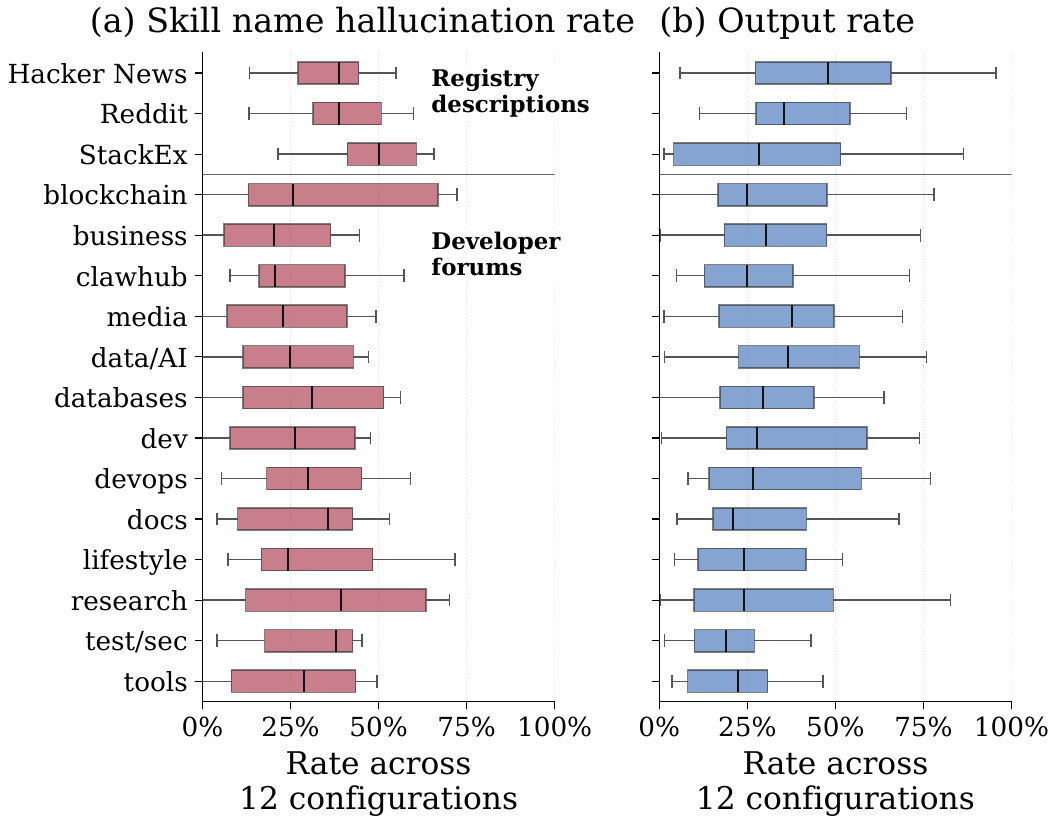}
  \caption{Hallucination rate by prompt source.}
  \label{fig:realsynth}
\end{figure}

This vulnerability is particularly pronounced under real-world usage conditions.
As \autoref{fig:realsynth} shows, forum queries pose a consistently higher risk
than synthesized skill description prompts. The median hallucination rate reaches
38.8\% on Hacker News and Reddit, and peaks at 50.1\% on Stack Exchange. Because
forum questions closely resemble the queries an attacker's victim would issue in
practice, this gap expands the real-world attack surface.

\begin{figure}[htbp]
  \centering
  \includegraphics[width=\columnwidth]{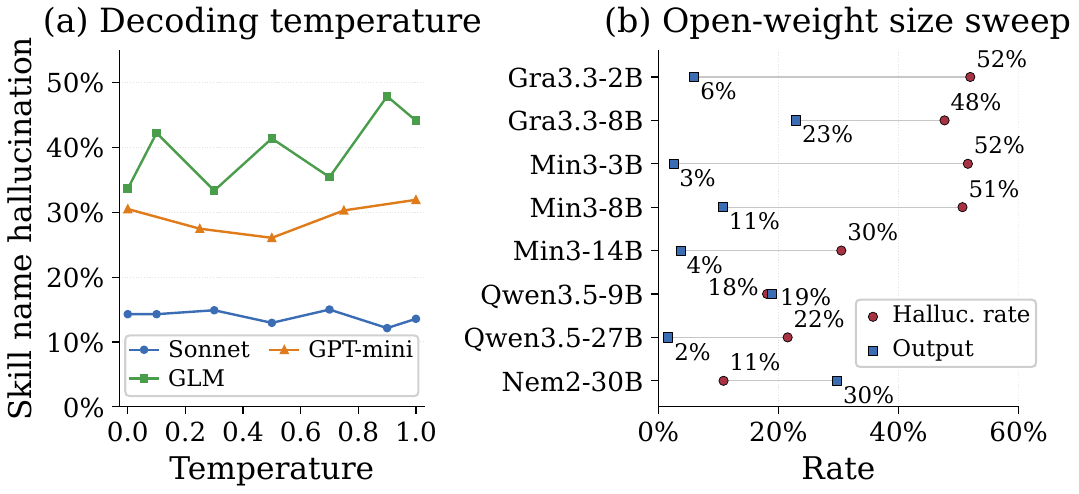}
  \caption{Hallucination rates across decoding temperatures and model sizes.}
  \label{fig:rq2_sensitivity}
\end{figure}

To assess whether this attack surface can be mitigated through hyperparameter tuning or model scaling, we evaluated various decoding temperatures and parameter sizes using prompt subsets of 4{,}000 prompts. \autoref{fig:rq2_sensitivity}(a) shows that changing the temperature for search-enabled LLMs from 0 to 1.0 produces little improvement. The per-prompt traces show surface-level variation without recovery. For the Retell AI troubleshooting skill, whose real name is \texttt{\seqsplit{retellai-common-errors}}, GLM-4.5 produced \texttt{\seqsplit{retellai-troubleshooting}}, \texttt{\seqsplit{retellai-troubleshooter}}, \texttt{\seqsplit{fix-retell-ai}}, and \texttt{\seqsplit{retellai-debug}} across temperatures, but never the real name. Similarly, for the CPU monitoring skill \texttt{\seqsplit{monitoring-cpu-usage}}, the same model alternated between \texttt{\seqsplit{cpu-performance-monitor}} and \texttt{\seqsplit{cpu-performance-analyzer}}. These examples illustrate that temperature mainly changes which plausible paraphrase is emitted, rather than forcing the model to verify the registered identifier.

\autoref{fig:rq2_sensitivity}(b) shows a similar pattern for model scaling across Granite~3.3, Ministral-3, Qwen~3.5, and Nemotron-Cascade-2. Larger variants sometimes reduce the aggregate rate, such as the drop in Ministral-3 from 51\% at 8B to 30\% at 14B, but they still produce semantically close false names. In the Ministral-3~14B run, the model mapped \texttt{\seqsplit{k8s-deploy}} to \texttt{\seqsplit{kubernetes-progressive-delivery}} and mapped \texttt{\seqsplit{rust-skill-creator}} to both \texttt{\seqsplit{crate-skill-generator}} and \texttt{\seqsplit{rustdoc-skill-generator}}. Qwen~3.5~27B similarly mapped \texttt{\seqsplit{exa-incident-runbook}} to \texttt{\seqsplit{execute-exa-incident-response-procedures}}, \texttt{\seqsplit{evidence-auditor}} to \texttt{\seqsplit{audit-claims-evidence}}, and \texttt{\seqsplit{openevidence-enterprise-rbac}} to \texttt{\seqsplit{openevidence-enterprise}}. Even the largest evaluated model, Nemotron-Cascade-2~30B, retained an 11\% hallucination rate over emitted names, corresponding to 390 hallucinated names among 3{,}588 recommendations on the shared 4{,}000-prompt subset. Thus, while scaling reduces the frequency of hallucinations across some model families, it fails to eliminate the underlying mechanism, as models continue to translate descriptions into plausible aliases similar to skills rather than recovering the exact published names.

\begin{findingbox}{Finding 2. The attack surface is structural, not a setting.}
No evaluated model, temperature, or size setting removes the problem. Search can
mask it, temperature mostly changes which hallucinated skill name appears, and larger models can
still hallucinate whenever they answer. The issue is built into the task, not one
special configuration.
\end{findingbox}

\subsection{RQ2: Targetability}
\label{sec:rq2}

The exploitability of this attack surface depends on predictability. To successfully execute a skill name squatting attack ahead of time, adversaries must anticipate which hallucinated names agents will generate before victims query them. Consequently, a hallucinated name becomes a high value target under two conditions: (1) high intra-model persistence, where a specific model reliably repeats the name to ensure consistent traps for victims, and (2) high inter-model convergence, where multiple models independently generate the same name to maximize the impact scope of a single registered payload. We evaluate both dimensions to quantify the targeting budget required for an attacker to achieve broad victim coverage.

\begin{figure}[htbp]
  \centering
  \includegraphics[width=\columnwidth]{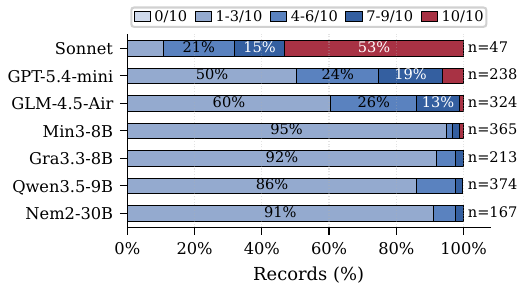}
  \caption{How often a model repeats hallucinated skill name across 10 runs.}
  \label{fig:rq3_persistence}
\end{figure}

Our experiment reveals high intra-model determinism for hallucinated names. By querying seven configurations 10 times each with 500 random prompts from RQ1, we define the dominant hallucination for each configuration and prompt as the confirmed hallucinated name that appears most often across the 10 runs, and then track how often it recurs. As shown in \autoref{fig:rq3_persistence}, Claude~Sonnet~4.6 exhibits striking persistence, repeating its dominant hallucination 7.8 times out of 10 on average, and returning the exact same name across all 10 runs in 53\% of cases. This stickiness stems from the model's adherence to instructions rather than local caching. Across all 10 trials, it generated \texttt{\seqsplit{citation-compass}} instead of the real \texttt{\seqsplit{citation-chasing-mapping}}, \texttt{\seqsplit{ideogram-monitoring-observability}} instead of \texttt{\seqsplit{ideogram-observability}}, and \texttt{\seqsplit{evernote-debug-bund}} instead of \texttt{\seqsplit{evernote-debug-bundle}}. While open-weight models are more volatile, repeating 1.5 to 2.0 times, highly persistent models allow attackers to reliably predict future outputs based on a single observed recommendation.

Beyond intra-model persistence, hallucinated names demonstrate significant inter-model convergence. Among the 1,728 prompt and model pairs with confirmed hallucinations, 219 dominant names were generated by at least two distinct models. For example, six independent models produced \texttt{classification-model-builder}, and five converged on names like \texttt{exa-incident-response}, \texttt{firebase-\\genkit} and \texttt{debugging-guidelines}. This indicates that models share underlying naming heuristics. When presented with a plausible capability, they predictably converge on identical hallucinated names. For an attacker, this shared behavior means a single hallucinated name can compromise users across entirely different LLM ecosystems.

A measurable sign of this shared naming prior is overlap with the broader developer software ecosystem. To test whether hallucinated skill names reflect neighboring package namespaces, we queried PyPI and npm for the 5{,}669 unique confirmed hallucinated names. As shown in \autoref{tab:rq2_cross_ecosystem}, 851 names, or 15.0\%, exactly match an existing package, even though the prompts requested agent skills rather than software packages. These matches are not valid skills. They indicate that models sometimes reuse familiar package or tool identifiers as names for skills. The overlap is concentrated in single configurations, but 126 names recur across two or more configurations, supporting the hypothesis that neighboring ecosystems shape predictable cross-model names.

\begin{table}[htbp]
  \centering
  \caption{Hallucinated skill names that match a PyPI or npm package}
  \label{tab:rq2_cross_ecosystem}
  \small
  \resizebox{\linewidth}{!}{
  \begin{tabular}{crrrc}
    \toprule
    \textbf{\#Configs} & \textbf{Records} & \textbf{Unique names} & \textbf{\% of total\textsuperscript{*}} & \textbf{PyPI / npm} \\
    \midrule
    1 & 725   & 725   & 12.8\%   & 349 / 376 \\
    2 & 170   & 85    & 1.5\%    & 44 / 41 \\
    3 & 90    & 30    & 0.5\%    & 19 / 11 \\
    4 & 32    & 8     & 0.1\%    & 5 / 3 \\
    5 & 5     & 1     & $<$0.1\% & 0 / 1 \\
    6 & 6     & 1     & $<$0.1\% & 0 / 1 \\
    8 & 8     & 1     & $<$0.1\% & 1 / 0 \\
    \midrule
    \textbf{Total} & \textbf{1{,}036} & \textbf{851} & \textbf{15.0\%} & \textbf{418 / 433} \\
    \bottomrule
  \end{tabular}
  }
  \parbox{\linewidth}{\scriptsize \textsuperscript{*}Percentage out of 5{,}669 unique, confirmed hallucinated names.}
\end{table}

\begin{figure}[htbp]
  \centering
  \includegraphics[width=\columnwidth]{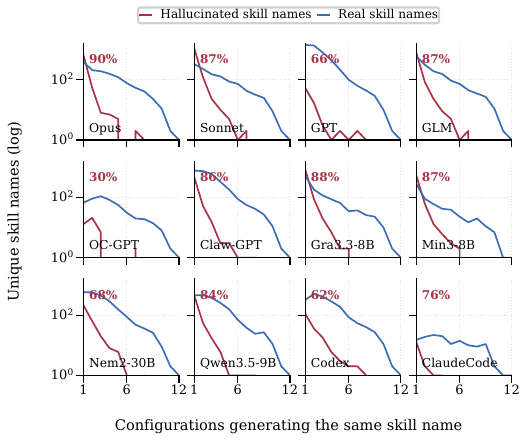}
  \caption{Cross-configuration reuse of hallucinated and real skill names.}
  \label{fig:rq4_overlap}
\end{figure}

While the absolute number of unique hallucinated names is vast and highly fragmented across configurations, the actual event frequency is heavily concentrated in shared names. \autoref{fig:rq4_overlap} illustrates the full set of names recovered from the 15,000-prompt corpus. Counted purely by distinct identifiers, 92.8\% of hallucinated names are exclusive to a single configuration. However, weighting these names by recommendation frequency reveals a different threat landscape. The 410 names that appear in at least two configurations account for 15.3\% of all 6,212 hallucination events, which is double their 7.2\% share by distinct count. Thus, most hallucinated names appear only once, but the names that recur are recommended often enough to become highly reusable targets.

\begin{figure}[htbp]
  \centering
  \includegraphics[width=\columnwidth]{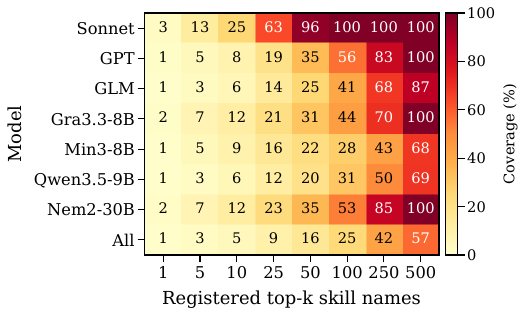}
  \caption{Coverage from registering top-ranked hallucinated skill names.}
  \label{fig:squat}
  \label{fig:targetability}
\end{figure}

This concentration allows attackers to achieve high impact with a minimal targeting budget. \autoref{fig:targetability} ranks hallucinated names by recurrence frequency to calculate the coverage an attacker would achieve by squatting the top-$k$ names. Each cell reports the percentage of hallucinated recommendations covered at that budget, and the ``All'' row pools events across the seven configurations in the persistence experiment. For Claude Sonnet, registering merely the top 25 names intercepts 63\% of all its hallucinated recommendations, and 50 names cover 96\%. Even aggregated across all seven configurations, squatting the top 500 names covers 57\% of all hallucinated recommendations. An attacker does not need to predict every rare hallucination, and a cheaply harvested dictionary of a few hundred predictable names sits exactly in the execution path of a majority of victims.

\begin{findingbox}{Finding 3: Hallucinations exhibit deterministic and cross-model patterns, enabling a highly efficient ahead-of-time skill name squatting attack.}
Rather than generating random noise, models draw from shared naming priors as they reliably repeat specific hallucinated names with a frequency of up to 7.8 out of 10 times, independently converge on identical names across different LLM families, and overlap with PyPI or npm package names for 15.0\% of unique hallucinated skill names. Due to this concentration, an attacker who squats the top 500 harvested names would cover 57\% of hallucinated recommendation events in our seven-configuration persistence experiment.
\end{findingbox}


\begin{casebox}{Case study: binding repeated names to installable skills.}
To verify the registration and installation path for predictable targets, we ran a benign proof-of-concept in an isolated Claude Code sandbox. We first selected \texttt{classification-model-builder}, a name our persistence experiment proved Claude Code reliably hallucinates for a specific prompt. We ethically created a public GitHub repository, added a benign \texttt{SKILL.md} file matching this name, and published a v1.0.0 release. We then fed the original targeted prompt to Claude Code. As expected, the agent recommended the hallucinated skill. We then instructed the agent to install this recommendation with a GitHub access token enabled, the agent automatically searched the platform, found our repository, and successfully installed the placeholder skill.

This confirms the delivery channel without using a malicious payload. Once an attacker registers a persistent hallucinated name, the agent workflow can turn a model mistake into an installed attacker-controlled skill.
\end{casebox}

\subsection{RQ3: Detectability}
\label{sec:rq3}

While RQ1 and RQ2 demonstrate the prevalence and predictability of this attack surface, an effective defense depends on whether defenders can identify hallucinated names prior to execution. We systematically evaluate two detection strategies: LLM self-auditing and typosquatting filters based on edit-distance analysis.

We first evaluate whether models can identify hallucinations in generated skill names. We asked seven models a yes/no question to check if a given skill name exists. Each model judged real skills against confirmed hallucinations under two conditions. The first is same-model testing for its own hallucinations. The second is cross-model testing for hallucinations pooled from the other six models. After removing unclear non-binary replies, we retain 3{,}707 of 4{,}448 model responses as valid binary judgments.

\begin{figure}[htbp]
  \centering
  \includegraphics[width=\columnwidth]{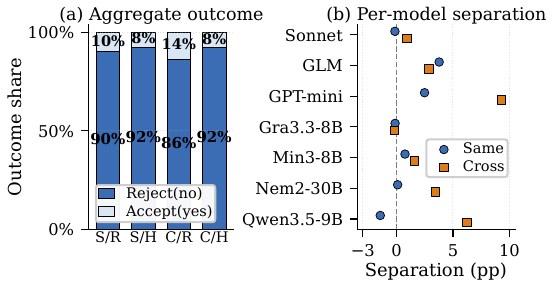}
  \caption{LLM self-detection. (S/C: same-/cross-model, R/H: real/hallucinated)}
  \label{fig:rq3_detection}
\end{figure}

\autoref{fig:rq3_detection} illustrates why this defense is fundamentally ineffective. As shown in \autoref{fig:rq3_detection}(a), the models overwhelmingly default to ``no'' and reject almost everything, specifically 90.6\% of names in the same-model condition and 88.8\% in the cross-model condition. While this extreme bias safely blocks 91\% of hallucinations, it destroys the agent's usefulness. The models recognize only 10.0\% of real skills among their own outputs and 13.6\% among those of other models. Even when a model does answer ``yes'' to its own generated names, it is still wrong 22.7\% of the time. Consequently, the per-model separation shown in \autoref{fig:rq3_detection}(b) yields an aggregate balanced accuracy of merely 51\% for same-model tests and 53\% for cross-model tests. Averaged across the two classes, the performance is no better than guessing.

The same-model and cross-model comparison helps explain this failure. A model performs worse at judging its own hallucinations than those of its peers because it uses the same internal logic to review the name as it did to generate it. GPT-5.4-mini is the only partial exception. It shows a less extreme rejection bias of 72.7\% and the highest real-skill recall of 28.0\%, yet it still fails to audit reliably. Ultimately, a model cannot filter a mistake it is built to make.

\begin{findingbox}{Finding 4. A model cannot audit an error it is built to make.}
Asked to distinguish real skill names from hallucinated ones, a model is no better than a coin flip,
with 51\% balanced accuracy on same-model tests and 53\% on cross-model tests. Same-model auditing also has lower overall accuracy than cross-model testing, 32.2\% vs 46.6\%, because models reject many real skills along with hallucinated ones. The self-auditing step inherits the same naming prior that produced the
mistake in the first place, so it cannot reliably catch the flaw.
\end{findingbox}

\begin{figure}[htbp]
  \centering
  \includegraphics[width=\columnwidth]{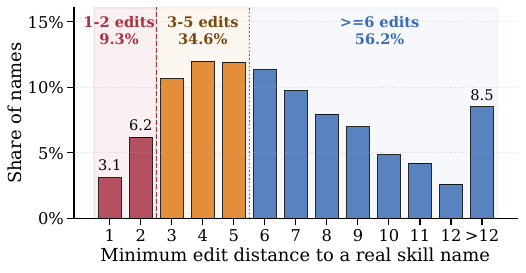}
  \caption{Distribution of Levenshtein distances of hallucinated/real skill names.}
  \label{fig:rq3_levenshtein}
\end{figure}

Traditional typosquatting defenses are inadequate because hallucinated names are novel semantic constructs rather than minor misspellings. Standard typosquatting filters typically flag variations within a Levenshtein distance of one or two from known real skills. However, comparing 5{,}669 confirmed hallucinated names against a catalog of 22{,}805 real skills reveals a mean edit distance of 6.85, as shown in \autoref{fig:rq3_levenshtein}. While 9.3\% of hallucinations represent near-misses, such as single substitutions, that such filters can catch, the vast majority, 56.1\% at $\ge6$ edits, bypass them entirely. These distant outliers are not typos, but rather long, newly synthesized capability descriptions, rendering structural similarity checks useless.

\begin{findingbox}{Finding 5. Hallucinated names are new phrases rather than typos.}
With a mean distance of 6.85 edits and a median of 6, hallucinated names are usually too far from any real skill for a typosquatting filter to catch. Only 9.3\% of the names sit close enough to look like ordinary misspellings. Therefore, edit-distance filters are largely ineffective. They catch only the few errors that resemble simple typos and completely miss the vast majority of newly invented names.
\end{findingbox}

\subsection{RQ4: Mitigation}
\label{sec:mitigation}

Finally, we investigate whether defenders can effectively mitigate this attack surface and evaluate the corresponding trade-offs regarding agent utility. To this end, we systematically evaluate four defensive strategies. Three operate within a single-model context, including RAG, self-refinement, and their combination. The fourth is a cross-model defense that requires consensus among multiple independent models before accepting a generated name. We applied the single-model defenses to the four open-weight models from RQ1 using a standardized subset of 2{,}000 prompts. \autoref{tab:mitigation} summarizes the resulting hallucination and no-output rates, where no output denotes cases producing no skill recommendation.

\begin{table}[htbp]
  \centering
  \caption{Skill name hallucination rate and no-output rate}
  \label{tab:mitigation}
  \resizebox{\linewidth}{!}{
  \begin{tabular}{lcccccccc}
    \toprule
    & \multicolumn{2}{c}{\textbf{Baseline}} & \multicolumn{2}{c}{\textbf{RAG}}
    & \multicolumn{2}{c}{\textbf{Self-refinement}} & \multicolumn{2}{c}{\textbf{RAG + self-refinement}} \\
    \textbf{Model} & Hall. & No output & Hall. & No output & Hall. & No output & Hall. & No output \\
    \midrule
    Ministral-3~8B           & 61.3\% & 81.9\% &  1.6\% & 74.2\% & 65.0\% & 99.3\% &  0.4\% & 86.9\% \\
    Granite~3.3~8B           & 46.4\% & 70.2\% & 10.0\% & 63.0\% & 38.3\% & 56.1\% &  9.8\% & 69.7\% \\
    Qwen~3.5~9B              & 32.5\% & 77.0\% &  0.9\% & 76.2\% & 11.4\% & 97.5\% &  0.7\% & 88.5\% \\
    Nemotron-Cascade-2~30B   & 22.8\% & 66.5\% &  0.6\% & 58.4\% & 18.0\% & 96.7\% &  0.1\% & 72.5\% \\
    \midrule
    \textbf{Mean}            & 40.8\% & 73.9\% &  3.2\% & 68.0\% & 33.2\% & 87.4\% &  2.7\% & 79.4\% \\
    \bottomrule
  \end{tabular}
  }
\end{table}

We first evaluate whether RAG can neutralize the attack surface by grounding models in real skill names. Using RAG~\cite{lewis2020rag}, we build a vector index over real skill names and their available descriptions with \texttt{all-MiniLM-L6-v2} sentence embeddings and the Chroma backend. For each user query, we embed the original prompt, retrieve the ten nearest skills by semantic similarity, format their names as a plain-text reference list, and prepend this list to the query under a system instruction that permits recommendations only from the retrieved names. We use ten retrieved skills to provide enough candidate coverage while keeping the context short. As shown in \autoref{tab:mitigation}, this defense yields the largest drop in hallucinations. For example, Qwen~3.5~9B falls from 32.5\% to 0.9\%, and Ministral-3~8B from 61.3\% to 1.6\%. On average, RAG cuts the hallucination rate by 37.6 percentage points. However, this defense is not free. It requires index maintenance, adds token costs, and creates retrieval-poisoning exposure~\cite{zou2025poisonedrag,chen2024agentpoison}.

We next test whether the detection failure observed in RQ3 carries over to an end-to-end mitigation pipeline. In the self-refinement setup~\cite{madaan2023selfrefine}, the model first generates candidate skill names and then receives its own output in a second turn, where it is instructed to remove any name it cannot verify. This internal validation reduces the average hallucination rate from 40.8\% to 33.2\%, but the reduction comes mainly from suppressing outputs rather than from reliable verification. The no-output rate rises from 73.9\% to 87.4\% on average, and becomes extreme for Qwen~3.5~9B and Nemotron-Cascade-2~30B, reaching 97.5\% and 96.7\%, respectively. Therefore, the RQ3 detection failure is not merely an artifact of binary yes/no testing: when used as an operational defense, self-refinement trades hallucinations for no output and still leaves a high hallucination rate among the few names it retains.

Given the isolated outcomes of RAG and self-refinement, we also examined whether combining them could offer compounding benefits. Running self-refinement after RAG only marginally changes the average hallucination rate, from 3.2\% under RAG to 2.7\%, while increasing the no-output rate from 68.0\% to 79.4\%. Once RAG restricts the model to real skills, the extra self-refinement pass mostly removes additional outputs instead of adding a meaningful safety benefit.

\begin{findingbox}{Finding 6. Giving real names helps. Self-refinement is unreliable.}
RAG reduces hallucination rates sharply, from a 40.8\% average down to 3.2\%. Self-refinement alone is much weaker: it lowers the rate mostly by returning no output, because the second pass repeats the same verification weakness as the first.
\end{findingbox}

\begin{figure}[htbp]
  \centering
  \includegraphics[width=\columnwidth]{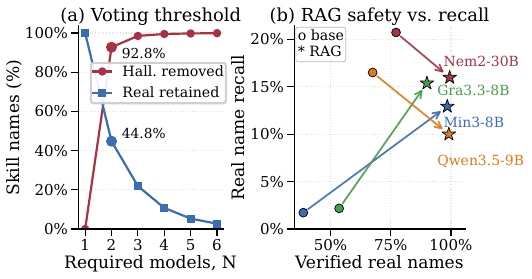}
  \caption{Mitigation trade-offs for voting and RAG.}
  \label{fig:rq4_mitigation_tradeoff}
\end{figure}

Shifting away from single-model strategies, we evaluated a cross-model voting defense. Building on our insight from RQ2 that hallucinations are usually specific to one model while real skills are shared, we hypothesize that requiring multiple models to agree could act as a natural filter. Indeed, as shown in \autoref{fig:rq4_mitigation_tradeoff}(a), if we accept a name only when two independent models produce it, corresponding to $N=2$, we eliminate 92.8\% of hallucinations while keeping 44.8\% of real skills. Raising the threshold to $N=3$ removes over 98\% of hallucinations, though real-skill preservation drops to 21.9\%. This sharp drop confirms that hallucinated names are isolated to single models. This approach requires no external database or model updates, needing only multiple independent agents.

While the aforementioned filters successfully suppress hallucinations, they introduce a severe penalty to the agent's overall utility. Because a defense that simply returns no output is not a practical fix, we also measure \emph{recall}, defined as the rate at which the agent successfully recommends the correct skill when a real one exists. Under RAG, safety and usefulness fundamentally trade off against each other. For the two weaker models, RAG improves recall by handing them correct answers they would not have guessed, with Ministral-3 rising from 1.7\% to 12.9\% and Granite~3.3 increasing from 2.2\% to 15.4\%. But for the two stronger models, recall still drops, as seen when Qwen~3.5 falls from 16.5\% to 10.0\% and Nemotron-Cascade-2 decreases from 20.7\% to 16.0\%, showing that retrieval constraints can suppress hallucinated names without reliably preserving the exact correct skill. Ultimately, even under RAG, no model recommends the correct skill more than 16.0\% of the time. \autoref{fig:rq4_mitigation_tradeoff}(b) shows that RAG shifts outputs toward verified real names without making real-name recall reliably high. The agent becomes safer mainly by returning fewer hallucinated names, not by becoming reliably correct.

\begin{findingbox}{Finding 7. The strongest defense buys safety with no output.}
Requiring two independent models to agree removes most hallucinations, but it also drops many real skills. Retrieval shows the same tradeoff, as the agent becomes safer mainly by answering less, not by naming the right skill reliably.
\end{findingbox}

\section{Discussion and Limitations}
\label{sec:discussion}

\noindent\textbf{Discussion.}
Our hallucination rates of 6.5\% to 62.0\% exceed the 5.2\% to 21.7\% reported for traditional packages~\cite{spracklen2025we}. This gap comes from a lack of knowledge about skills. The Agent Skills standard appeared in December 2025~\cite{anthropic_skills_2025}, after the training cutoff of most models we test, so models often fall back on familiar package names from ecosystems such as npm~\cite{sonatype2026ssc} and PyPI~\cite{librariesio2026}. As a result, a name like \texttt{eslint-config-prettier} is returned as a skill (Section~\ref{sec:rq2}), making the same hallucinated name useful for attacks across multiple ecosystems.

This explains the inadequacy of simple defenses. Web search verifies a tool's existence, but not its validity as a skill. Self-refinement has the same problem because the second pass evaluates the answer using the same prior that produced the hallucination. RQ4 shows the cost of suppressing these names. Even the agent with the strongest defense returns the correct skill at most one time in six, while several configurations reduce hallucination mainly by returning no output. This trades false positives for false negatives, producing safer recommendations but fewer useful ones. The core challenge is therefore not silence, but retrieval of the right skill without admitting hallucinated skill names.

This trade-off also suggests that skill registries need defenses that are more explicit than package registries. Because a skill is primarily a natural-language instruction bundle, artifact-level checks provide weaker signals than they do for ordinary packages. A practical defense should therefore treat recommendation as a registry lookup. The agent should retrieve from an authenticated skill index and refuse names that cannot be resolved to a concrete \texttt{SKILL.md}. Namespace reservation for recurring hallucinated names may reduce ahead-of-time squatting, but it cannot replace exact lookup because our results show a long tail of model-specific names.

\noindent\textbf{Limitations.}
Our study measures the attack \emph{precondition}, in which agents hallucinate reusable skill names and defenses often miss them. We also verify the registration and installation path with a benign placeholder skill in an isolated sandbox, but we omit a full malicious compromise. Ethically, we did not register malicious skills, attack any public registry, or expose real users to the workflow. The exploitation step itself follows a known mechanism. Once a poisoned \texttt{SKILL.md} is installed, it enters the agent context and can act as instruction level prompt injection, as shown in prior work~\cite{greshake2023_indirect,liu2024_prompt_injection, guo2026malskillbench}. We establish this delivery channel is plausible because attackers can predict attractive names and register them in advance. Measuring whether real users install such recommendations requires a controlled user study. Finally, our recall analysis covers only the four open weight models used for mitigation, so we defer the full false-negative study across all 12 configurations.

\section{Related Work}

\subsection{Supply Chain Attacks and Package Hallucination}
While traditional supply chain attacks~\cite{ohm2020backstabber, duan2021measuring, ladisa2023sok, neupane2023typosquatting} exploit mistyped legitimate packages, LLMs introduce novel threats via hallucinated recommendations. Following discoveries of package hallucinations~\cite{spracklen2025we}, recent research categorized code generation errors~\cite{zhang2025codehallucination} and proposed API-level mitigations~\cite{jain2025apihallucination}. However, extending these insights to the Agent Skills ecosystem demands a fundamental threat model shift. Unlike mature code registries supporting artifact comparison against established references, skill hallucinations lack legitimate counterparts. Furthermore, skills execute natural language instead of compiled code, rendering static analysis ineffective.

Our work is closest to package hallucination, but differs in both the object being named and the user's trust path. A hallucinated package must usually survive dependency resolution, build tooling, and code review before it affects an application. A hallucinated skill can be installed because the same agent that recommends it also helps fetch and activate it. The payload is also not limited to executable source code. Adversarial instructions can be embedded directly in a \texttt{SKILL.md}, where they become part of the agent's future context. We therefore study not only whether names are false, but whether false names are persistent, reusable, and attractive enough for attackers to register before victims ask for them.

\subsection{Agent Security and Prompt Injection}
Installed malicious skills inherently threaten the execution integrity of LLM agents. Recent studies have surveyed widespread vulnerabilities in agent applications~\cite{he2025agentsecurity, shen2025securitydebt, ruan2023toolemu, yuan2024rjudge, andriushchenko2025agentharm}, demonstrating how malicious modules or online interactions amplify risks like data exfiltration and impersonation~\cite{dong2025philosopher, kim2025llmonline}. A critical consequence of such compromises is indirect prompt injection, extensively benchmarked in recent literature~\cite{greshake2023_indirect, zhan2024injecagent, liu2024_prompt_injection, debenedetti2024agentdojo, shi2025toolhijacker}. Contemporary defenses utilize structured query separation~\cite{chen2025_struq}, preference optimization~\cite{chen2025secalign}, execution isolation~\cite{wu2025isolategpt}, and guardrail agents~\cite{xiang2025guardagent} to sanitize adversarial inputs during inference. Yet, these methods implicitly assume the malicious payload arrives alongside the user query. Skill name hallucination circumvents these boundaries entirely. By tricking the user into installing a poisoned skill, the adversary injects adversarial instructions directly into the persistent configuration of the agent, bypassing standard alignment and runtime protections~\cite{qi2024finetuning, zheng2024fewshot, liu2025datasentinel}. This makes installation a persistent and reusable trust boundary.

\subsection{Threats in the Agent Skills Ecosystem}
Consequently, the Agent Skills ecosystem presents a unique and largely unaddressed attack surface. Liu et al.~\cite{liu2026skillsinthewild} identify widespread vulnerabilities within published skills, while Qu et al.~\cite{qu2026supplychainpoisoning} demonstrate that malicious logic embedded in skill documentation can successfully manipulate agent behavior. Both works assume the attacker compromises existing skills or publishes payloads hoping for manual discovery. Our research complements these studies by formalizing vulnerabilities at the recommendation phase. This shifts the attack earlier than prior work considers. We demonstrate that attackers do not need to wait for organic discovery, but rather can weaponize the persistent hallucinations of LLMs in advance to actively drive victims toward malicious skills.

\section{Conclusion}


We studied skill name hallucination as a supply-chain attack across 12
configurations and 15{,}000 prompts. Every one hallucinated, at 6.5--62.0\%
overall and 43.1\% on real questions. The names are targetable: models repeat a
name up to 7.8 in 10, 410 shared names cover 15.3\% of hallucinations, and 851 are
real PyPI or npm packages. They also evade detection: self-checks match chance,
and names sit a median of six edits from any real skill. RAG cuts hallucination by
37.6 percentage points, yet the best-defended agent recommends the right skill at most one in
six. It is therefore a real, low-cost attack surface, and closing it requires
registry-level reservation and verified recommendation, not tuning alone.

\section{Data Availability}
\label{sec:data_availability}

 We release all prompts, skills, and responses at \url{https://figshare.com/s/4e408bce9b2f64224f79}.

\bibliographystyle{ACM-Reference-Format}
\bibliography{references}

\end{document}